\def\breakon{\end{multicols}\widetext\vspace{.5cm}
\noindent\rule{.48\linewidth}{.3mm}\rule{.3mm}{.5cm}\vspace{.5cm}}
\def\breakoff{\vspace{.5cm}
\noindent
\rule{.52\linewidth}{.0mm}\rule[-.47cm]{.3mm}{.5cm}\rule{.48\linewidth}{.3mm}
\vspace{.5cm}
\begin{multicols}{2}
\narrowtext}
\documentstyle[aps,epsf,prb,multicol]{revtex}
\begin{document}

\makeatletter
\renewenvironment{table}
  {\let\@capwidth\linewidth\def\@captype{table}}
  {}

\renewenvironment{figure}
  {\let\@capwidth\linewidth\def\@captype{figure}}
  {}
\makeatother

\title{Recurrent Variational Approach to the Two-Leg Hubbard Ladder}
\author{Eugene H. Kim}
\address{Department of Physics, University of California, 
         Santa Barbara, California 93106-9530}
\author{Germ\'{a}n Sierra}
\address{Instituto de Matem\'{a}ticas y F\'{i}sica Fundamental,
         C.S.I.C., 28006 Madrid, Spain}
\author{Daniel Duffy}
\address{Department of Physics, University of California, 
         Santa Barbara, California 93106-9530}
\maketitle

\begin{abstract}
We applied the Recurrent Variational Approach to the two-leg
Hubbard ladder.  At half-filling, our variational Ansatz was 
a generalization of the resonating valence bond state.  At 
finite doping, hole pairs were allowed to move in the resonating 
valence bond background.  The results obtained by the Recurrent
Variational Approach were compared with results from 
Density Matrix Renormalization Group.
\end{abstract}

\vspace{.15in}
\begin{multicols}{2}


\section{Introduction}
In the hope to get a better understanding of strongly interacting
systems, there has been considerable interest in ladder systems. \cite{rice}
These ladder systems have proven to be a theoretical
wonderland, both analytically \cite{analytic} and numerically. 
\cite{numerical}  However, much of the analytic work done on ladders has
been in weak coupling (or perturbatively in some parameter), namely 
because there are very few analytic methods at strong coupling.  
Exact diagonalization, Monte Carlo, and Density Matrix Renormalization 
Group methods have been the primary tools for studying these systems
at strong coupling.  Each of these methods has both strengths and
weaknesses when considering the lattice sizes, temperatures, and
couplings on can consider.

With the ability to fabricate these materials, \cite{fabricate} 
ladders are not just a theoretical playground.  For example, 
in $(VO)_2P_2O_7$, there are well separated two-leg ladders
composed of $VO_4$. \cite{vo4}  Also the cuprate-like material 
$SrCu_2O_3$ consists of weakly coupled $CuO_2$ two-leg ladders,
\cite{cuo2} and the material $Sr_2Cu_3O_5$ consists of weakly coupled 
$CuO_2$ three-leg ladders. \cite{cuo3}

Recently, a powerful analytic method was developed to deal
with strongly coupled quasi-one dimensional systems ---
the Recurrent Variational Approach (RVA). \cite{sierra1,sierra2} 
This method is similar in spirit to Wilson's Numerical 
Renormalization Group \cite{wilson} and White's Density Matrix 
Renormalization Group (DMRG). \cite{white}  The key idea in all of these 
methods is to build up the system by adding on sites at the boundary.  
However, the real power of the RVA is that, though analytic, the physics
of the problem is taken into account in an unbiased way and elucidated
quite clearly.  For example, the importance of different configurations 
in the ground state wave function is determined without any outside 
assumptions, and the physics of these configurations is made clear.

In this work, we apply the RVA to the two-leg Hubbard ladder
at strong coupling and small dopings.  Though a considerable
amount of work has been done on the Hubbard ladder and many
of its properties are known, we are unaware of any work which
has put this information together and constructed a ground state
wave function.  Our goal in this work is to provide
a simple physical picture of what the ground state might look like
and to go ahead and construct a ground state wave function.

The Hamiltonian of the two-leg Hubbard ladder is given by
\begin{eqnarray}
 H & = & -t \sum_{ {\bf i},s } 
       \left( c^{\dagger}_{{\bf i},s} 
              c^{\phantom \dagger}_{{\bf i + \hat{x}},s} 
              + h. c. \right)
         -t_{\perp} \sum_{ {\bf i},s } 
       \left( c^{\dagger}_{{\bf i},s} 
	      c^{\phantom \dagger}_{{\bf i + \hat{y}},s} 
              + h. c. \right)  \nonumber \\
   & & \hspace{.75in} +U \sum_{{\bf i}} n_{{\bf i}, \uparrow} 
                                        n_{{\bf i}, \downarrow}  \, ,
\end{eqnarray}
where $c^{\dagger}_{{\bf i},s}$ creates a fermion at site {\bf i} with
spin $s$, 
$n_{{\bf i},s} = c^{\dagger}_{{\bf i},s} c^{\phantom \dagger}_{{\bf i},s}$, 
$t$ is the hopping matrix element along the chain, $t_{\perp}$
is the hopping matrix element perpendicular to the chain (i.e., along the
rung), and $U$ is the on-site Coulomb repulsion.  Site {\bf i} has 
coordinates $(x,y)$ with $1 \leq x \leq N$ and $y=1,2$.
It will also be convenient to introduce the following two operators:
\begin{equation}
 \Delta_{{\bf i},{\bf j}}^{\dagger} = 
 c^{\dagger}_{\bf i,\uparrow} c^{\dagger}_{\bf j,\downarrow} +
 c^{\dagger}_{\bf j,\uparrow} c^{\dagger}_{\bf i,\downarrow} \, ,
\label{singlet}
\end{equation}
which creates a singlet across sites {\bf i} and {\bf j}, and
\begin{equation}
 D_{{\bf i}} = 
 c^{\dagger}_{\bf i,\uparrow} c^{\dagger}_{\bf i,\downarrow} \, , 
\label{double}
\end{equation}
which creates a doubly occupied site.  

The rest of the paper is organized as follows.  In Sec.~II we consider
the half-filled case.  In Sec.~III we consider the Hubbard ladder at
small dopings.  In Sec.~IV we present our results for the ground state
energies and compare with DMRG.  Finally, in Sec.~V we summarize our 
results and present some concluding remarks.


\section{Half-Filled Hubbard Ladder}

We begin with the half-filled ladder.  What are the ingredients 
necessary to construct a wave function which captures the physics 
of the half-filled Hubbard ladder?  A trail of clues has been 
laid by previous works.  First of all, we know that at 
strong coupling the half-filled   \newpage

\begin{figure}
\epsfxsize= 3.25in
\centerline{\epsfbox{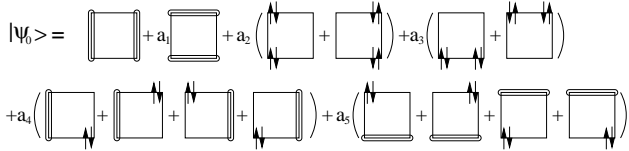}}
\caption{Ground state wave function for the 2x2 half-filled plaquette.
A circled link represents a singlet bond.  (See text for a full explanation
of each state.)}
\label{fig:2x2half}
\end{figure}
\vspace{.25in}

\noindent  Hubbard model is equivalent to
the Heisenberg model.  Secondly, it is known that the two-leg
Heisenberg ladder (and Hubbard ladder) have a spin gap and
short range spin correlations.  Furthermore, it has been shown 
that the resonating valence bond (RVB) state captures the essential
physics of the two-leg Heisenberg ladder.\cite{RVB,sierra1}
Hence, the RVB picture should capture the essential physics
of the half-filled Hubbard ladder at strong coupling. Thus
we propose a generalization of the RVB state as our variational ansatz.

A key property of the RVB state is that it is the exact
solution to the $2 \times 2$ Heisenberg plaquette.  Therefore, we
base our generalized RVB state on the exact solution to the
$2 \times 2$ Hubbard plaquette.  Since the $2 \times 2$ plaquette 
will serve as the basis for our ansatz, we discuss it in detail below. 


\subsection{The $2 \times 2$ Plaquette}
The ground state of the $2 \times 2$ plaquette is given by
\begin{eqnarray}
 \mid \psi_0 \rangle \  & = & \ \mid \varphi_0 \rangle  +  
                     a_1~\mid \varphi_1 \rangle  + 
		     a_2~\mid \varphi_2 \rangle  +
		     a_3~\mid \varphi_3 \rangle 
 \nonumber \\
               & + & a_4~\mid \varphi_4 \rangle  +
		     a_5~\mid \varphi_5 \rangle
\end{eqnarray}
where 
\begin{eqnarray}
 \mid \varphi_0 \rangle & = &
     \Delta^{\dagger}_{(1,1),(1,2)} \Delta^{\dagger}_{(2,1),(2,2)}
     \mid 0 \rangle \, , 
 \nonumber \\
 \mid \varphi_1 \rangle & = &
     \Delta^{\dagger}_{(1,1),(2,1)} \Delta^{\dagger}_{(1,2),(2,2)}
     \mid 0 \rangle \, ,
 \nonumber \\
 \mid \varphi_2 \rangle & = &
     D^{\dagger}_{(1,1)} D^{\dagger}_{(1,2)} \mid 0 \rangle +
     D^{\dagger}_{(2,1)} D^{\dagger}_{(2,2)} \mid 0 \rangle \, ,
 \nonumber \\
 \mid \varphi_3 \rangle & = & 
     D^{\dagger}_{(1,1)} D^{\dagger}_{(2,1)} \mid 0 \rangle +
     D^{\dagger}_{(1,2)} D^{\dagger}_{(2,2)} \mid 0 \rangle \, ,
 \nonumber \\
 \mid \varphi_4 \rangle & = &
     \Delta^{\dagger}_{(1,1),(1,2)} D^{\dagger}_{(2,1)} \mid 0 \rangle +
     \Delta^{\dagger}_{(1,1),(1,2)} D^{\dagger}_{(2,2)} \mid 0 \rangle  
 \nonumber \\
 & + &  \Delta^{\dagger}_{(2,1),(2,2)} D^{\dagger}_{(1,1)} \mid 0 \rangle +
     \Delta^{\dagger}_{(2,1),(2,2)} D^{\dagger}_{(1,2)} \mid 0 \rangle \, , 
 \nonumber \\
 \mid \varphi_5 \rangle & = &
     \Delta^{\dagger}_{(1,1),(2,1)} D^{\dagger}_{(1,2)} \mid 0 \rangle +
     \Delta^{\dagger}_{(1,1),(2,1)} D^{\dagger}_{(2,2)} \mid 0 \rangle  
 \nonumber \\
 & + &  \Delta^{\dagger}_{(1,2),(2,2)} D^{\dagger}_{(1,1)} \mid 0 \rangle +
     \Delta^{\dagger}_{(1,2),(2,2)} D^{\dagger}_{(2,1)} \mid 0 \rangle \, .
\end{eqnarray}
 $\mid \psi_0 \rangle$ is shown schematically in Fig.~\ref{fig:2x2half}.

We will be mainly interested in the case $t=t_{\perp}$.  In 
Table \ref{table:2x2half} 
we list the values of the parameters for several values of $U$ 
(with $t=t_{\perp} = 1$).

\vspace{.25in}
\begin{figure}
\epsfxsize= 3.1in
\centerline{\epsfbox{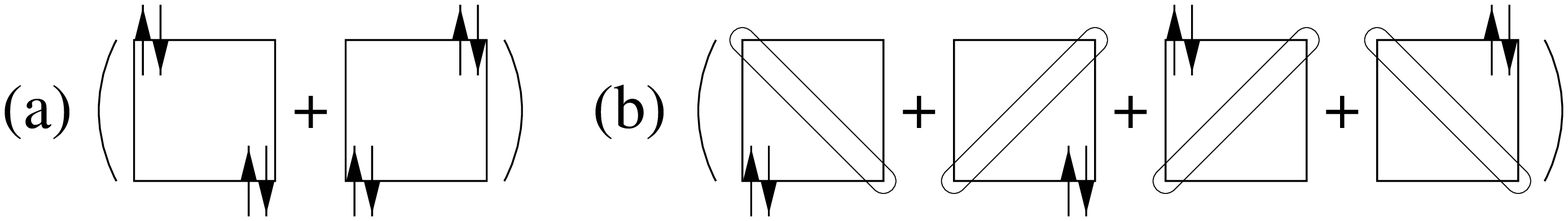}}
\caption{States which the $B_2$ representation prevents from 
appearing in the ground state wave function of the half-filled 
plaquette.}
\label{fig:badstates}
\end{figure}
\vspace{.2in}

\begin{table}
\caption{Values of the parameters for the half-filled 2x2 plaquette 
(with $t = t_{\perp} = 1$) which give the exact groundstate.}
\begin{center}
\begin{tabular}{|c|c|c|c|c|c|c|c|c|} 
\hline 
$U$ &  $a_1$ & $a_3$ & $a_5$ & $a_2$ & $a_4$ \\  \hline \hline

 $8$  & $-1.0$ & $-0.0762$ & $-0.3306$ & $-a_3$ & $-a_5$  \\ \hline 

 $16$ & $-1.0$ & $-0.0221$ & $-0.1807$ & $-a_3$ & $-a_5$  \\ \hline 

 $24$ & $-1.0$ & $-0.0101$ & $-0.1229$ & $-a_3$ & $-a_5$  \\ \hline 
\end{tabular}
\end{center}
\label{table:2x2half}
\end{table}
\vspace{.2in}

Notice that the solution to the 2x2 plaquette (with $t=t_{\perp}$)
has $D_4$ symmetry. ($D_4$ is the
symmetry group of the square.)  However, the ground state
does {\it not} transform in the scalar representation of $D_4$; it
transforms in the $B_2$ representation of $D_4$. ($B_2$ is the 
one-dimensional representation which changes sign upon rotation 
by $90^o$ and reflection about the diagonals.)  $B_2$ coincides
with the standard $d_{x^2-y^2}$ symmetry.  
The $B_2$ representation is what forbids some configurations, as 
those shown in Fig.~\ref{fig:badstates}, from appearing 
in the ground state wave function.

A few more words are in order about the ground state wave function.
Notice that $a_1 = -1$ (i.e., the weight of the horizontal singlets
is equal to (minus) the weight of the vertical singlets.)  This is
the RVB mechanism -- the system lowers its energy by resonating 
between horizontal and vertical singlets; $a_1$ is the 
``RVB parameter''.  Also, notice that 
$\mid a_2 \mid, \mid a_3 \mid = {\cal O}(a_4^2)$.  
This will play a role in constructing the Ansatz for the
half-filled ladder.


\subsection{The Ladder}
\begin{figure}
\epsfxsize= 2.9in
\centerline{\epsfbox{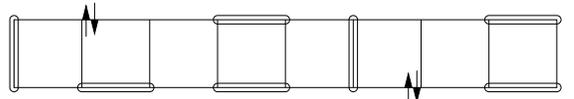}}
\caption{Typical configuration appearing in the ground state 
wave function of the half-filled ladder.}
\label{fig:typicalhalf}
\end{figure}

Using the configurations of the $2 \times 2$ plaquette as the basis 
of our Ansatz for the ladder, a typical configuration for the ladder
is shown in Fig.~\ref{fig:typicalhalf}.  
The ground state will be a superposition of all possible configurations
of the type shown in Fig.~\ref{fig:typicalhalf}.  Therefore, it seems
like working with this kind of state will be a formidable task.
Fortunately, the RVA gives us a straightforward way of dealing with
such a state -- generate it {\it recursively}.
Specifically, the RVA builds the ground state of a ladder with 
$N+\nu$ rungs using the knowledge of the ground
states of a ladder with $N, N+1, \ldots , N+\nu  -1$ rungs.  
This is achieved by recursion relations
which express the ground state $\mid N+\nu \rangle$ in terms
of the ground states 
$\{\mid N+i \rangle \}$ with $i=0, \ldots, \nu-1$.\cite{sierra1}

Using these ingredients, we consider the following ansatz for the
half-filled Hubbard ladder which is shown

\vspace{.5in}
\vspace{.2in}

\begin{figure}
\epsfxsize= 3.2in
\epsfysize= 2.0in
\centerline{\epsfbox{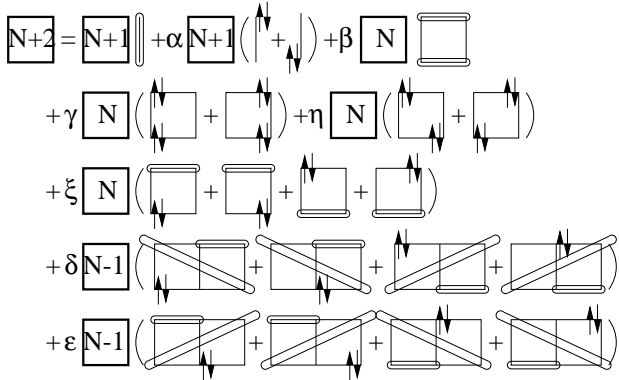}}
\caption{The RVA ansatz for the half-filled Hubbard ladder.  Note that
links or sites connected with a circle represent a singlet bond.}
\label{fig:goodstates}
\end{figure}
\vspace{.2in}

\begin{eqnarray}
 \mid N+2 \rangle\ &  = & \, \mid \phi_0 \rangle_{N+2} \mid N+1 \rangle 
                  + \alpha \mid \phi_1 \rangle_{N+2} \mid N+1 \rangle
 \nonumber \\
                 & + & \, \beta \mid \phi_2 \rangle_{N+1,N+2} \mid N \rangle
	          + \gamma \mid \phi_3 \rangle_{N+1,N+2} \mid N \rangle
 \nonumber \\
	         & + & \, \xi \mid \phi_4 \rangle_{N+1,N+2} \mid N \rangle 
	          + \eta \mid \phi_5 \rangle_{N+1,N+2} \mid N \rangle
 \nonumber \\
   & + & \, \delta \mid \phi_6 \rangle_{N,N+1,N+2} \mid N-1 \rangle
 \nonumber \\
   & + & \, \varepsilon \mid \phi_7 \rangle_{N,N+1,N+2} \mid N-1 \rangle  
\label{halffilledansatz}
\end{eqnarray}
schematically in Fig.~\ref{fig:goodstates}. \cite{tensorproduct}  
For completeness, the states in terms of the operators in
Eqs.~\ref{singlet} and \ref{double} are given in Appendix A.

\begin{figure}
\epsfxsize=2.5in
\centerline{\epsfbox{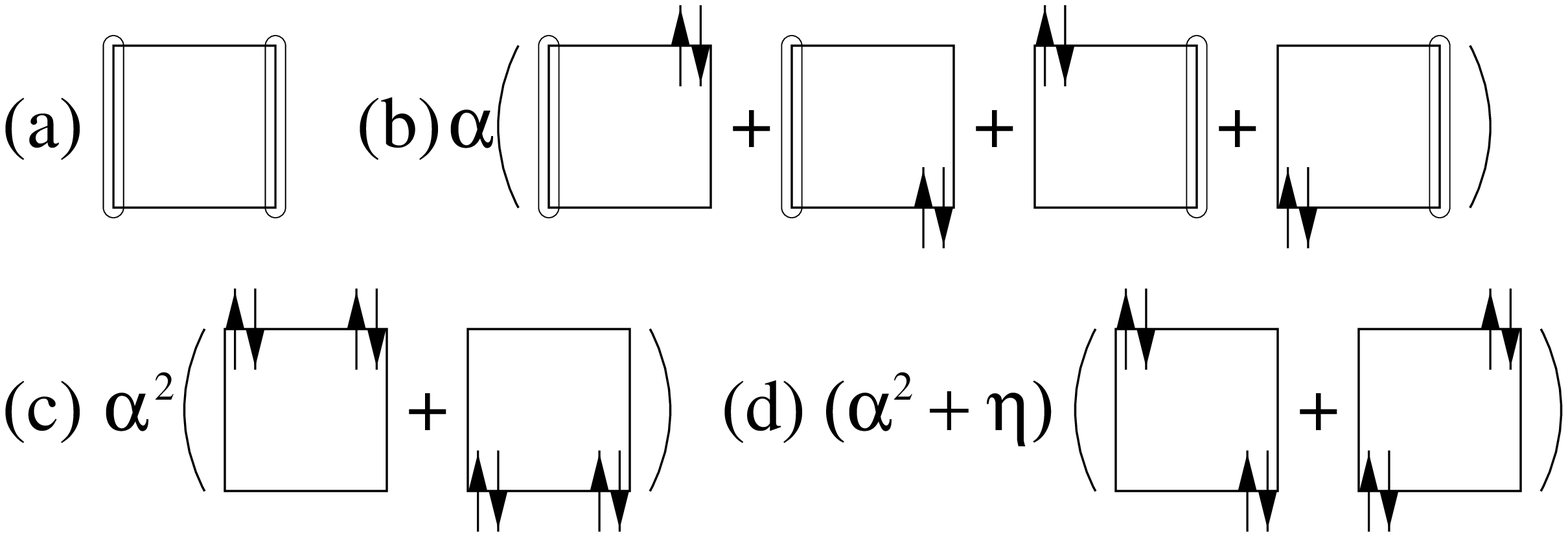}}
\caption{States generated by $\mid \phi_0 \rangle$ and 
$\mid \phi_1 \rangle$ (and also $\mid \phi_5 \rangle$)
by running the recursion relations once.}
\label{fig:generatehalf}
\end{figure}

A few words are in order about the configurations in our Ansatz.
i) $\beta$, the weight of the horizontal bond (relative to the vertical
bond), is the ``RVB parameter''.  For a true RVB state, we would
have $\beta =-1$.  Although we do not expect $\beta = -1$, from work
on the Heisenberg model the variational parameters were shown to 
evolve smoothly with system size; \cite{sierra1}  we expect
$\mid \beta \mid = {\cal O}(1)$.   
ii)  Suppose we iterate the recursion relations once.  Then 
 $\mid \phi_0 \rangle$ and $\mid \phi_1 \rangle$ (and also 
 $\mid \phi_5 \rangle$) generate the 
terms shown in Fig.~\ref{fig:generatehalf}.
In the ground state of the 2x2 plaquette, the states in 
Fig.~\ref{fig:generatehalf}(b) have weight ${\cal O}(\alpha)$, 
the states in Fig.~\ref{fig:generatehalf}(c) have weight 
${\cal O}(\alpha^2)$, and the states in Fig.~\ref{fig:generatehalf}(d) 
do not appear.  Since we expect the parameters to evolve smoothly, 
\cite{sierra1} we expect $\eta \approx -\alpha^2$.  
($\eta$ is behaving as a ``counter-term'';
it's job is to subtract off the $\alpha^2$ contribution from 
$\mid \phi_1 \rangle$.)  iii) Even though we no 
longer have $D_4$ symmetry, initial calculations showed that the states 
in Fig.~\ref{fig:badstates}(b) do not appear in the ground 
state of the Hubbard ladder.  (This is another indication that the 
parameters evolve smoothly from the 2x2 case to the ladder.)
Therefore, we do not consider them in what follows.  
iv) Since the configurations  $\mid \phi_6 \rangle_{N,N+1,N+2}$ and 
$\mid \phi_7 \rangle_{N,N+1,N+2}$ appear as intermediate states for the

\vspace{.25in}
\vspace{.25in}

\begin{figure}
\epsfxsize=2.25in
\centerline{\epsfbox{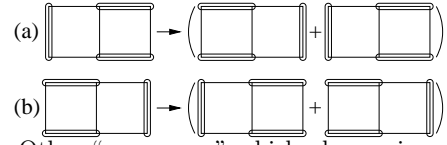}}
\caption{Other ``resonances'' which play an important role in 
the RVB picture.}
\label{fig:resonatenew}
\end{figure}
\vspace{.25in}

\noindent
resonances shown in Fig.~\ref{fig:resonatenew}, it is
necessary to include these states in the Ansatz to give us an RVB
state.  This can easily be seen by comparing the weights of the 
coefficients as shown in Table \ref{table:halfladder} (see below).

\vspace{.2in}
In order to compute the values of the coefficients, we treat them as 
variational parameters and minimize the ground state energy with 
respect to these parameters.  The ground state energy and other 
quantities appear as recursion relations.  It will be useful to define 
\begin{eqnarray}
 E_N \ &  = & \ \langle N \mid H_N \mid N \rangle \, ,  \nonumber \\
 D_N \ &  = & \ \langle N-1 \mid_N \langle \phi_0 \mid 
                 H_N \mid N \rangle \, ,  \nonumber \\
 C_N \ &  = & \ \langle N-1 \mid_N \langle \phi_1 \mid 
                 H_N \mid N \rangle \, ,   \nonumber \\
 Z_N \ &  = & \ \langle N \mid N \rangle \, ,   \nonumber \\
 Y_N \ &  = & \ \langle N-1 \mid_N \langle \phi_0 \mid 
                 N \rangle \, ,   \nonumber \\
 X_N \ &  = & \ \langle N-1 \mid_N \langle \phi_1 \mid 
                 N \rangle \, . 
\end{eqnarray}
They are supplemented by the initial conditions 
\begin{eqnarray}
 & & Z_0 = 1, Y_0 = 0, X_0 = 0, \nonumber \\
 & & E_0 = 0, D_0 = 0, C_0 = 0 \, .
\end{eqnarray}

To determine the values for the variational parameters for a given
(finite) value of $N$, we iterate the recursion relations and
minimize the quantity $E_N / Z_N$ numerically.
The actual recursion relations are quite unwieldly; we have 
relegated them, as well as their derivation, to Appendix A.

The values of the variational parameters for various values 
of $U$ (with $t = t_{\perp} = 1$) are shown in 
Table \ref{table:halfladder}.  
The results were obtained on a $2 \times 32$ ladder.
Notice that i) $\mid \beta \mid = {\cal O}(1)$, and we have 
produced an RVB state.  ii) $\mid \eta \mid = {\cal O}(\alpha^2)$, 
and $\eta$ is indeed behaving as a counterterm.  
iii) $\mid \delta \mid, \mid \varepsilon \mid \approx \alpha/3$.  
Therefore, these configurations are non-negligible, suggesting that 
the resonances shown in Fig.~\ref{fig:resonatenew} are important to 
the RVB picture.


\section{The Doped Hubbard Ladder}
Now, we consider the doped Hubbard ladder.  In Ref.~9 it was shown
that hole pairs moving through an RVB

\vspace{.25in}
\begin{table}
\caption{Values of the variational parameters for a $2 \times 32$
half-filled ladder with $t = t_{\perp} = 1$. }
\begin{center}
\begin{tabular}{|c|c|c|c|c|c|c|c|c|} 
\hline 
$U$ &  $\alpha$ & $\beta$ & $\gamma$ &
$\xi$ & $\eta$ & $\delta$ & $\varepsilon$\\  \hline \hline

 $8$ & $.3296$ & $-.8710$ & 
 $-.0782$ & $-.2877$ & $-.0938$ &
 $.1031$ & $.1031$  \\ \hline 

 $16$ & $.1848$ & $-.8800$ & 
 $-.0243$ & $-.1606$ & $-.0299$ &
 $.0639$ & $.0639$  \\ \hline 

 $24$ & $.1265$ & $-.8817$ & 
 $-.0113$ & $-.1097$ & $-.0142$ &
 $.0451$ & $.0451$  \\ \hline 
\end{tabular}
\end{center}
\label{table:halfladder}
\end{table}

\begin{figure}
\epsfxsize= 2.6in
\centerline{\epsfbox{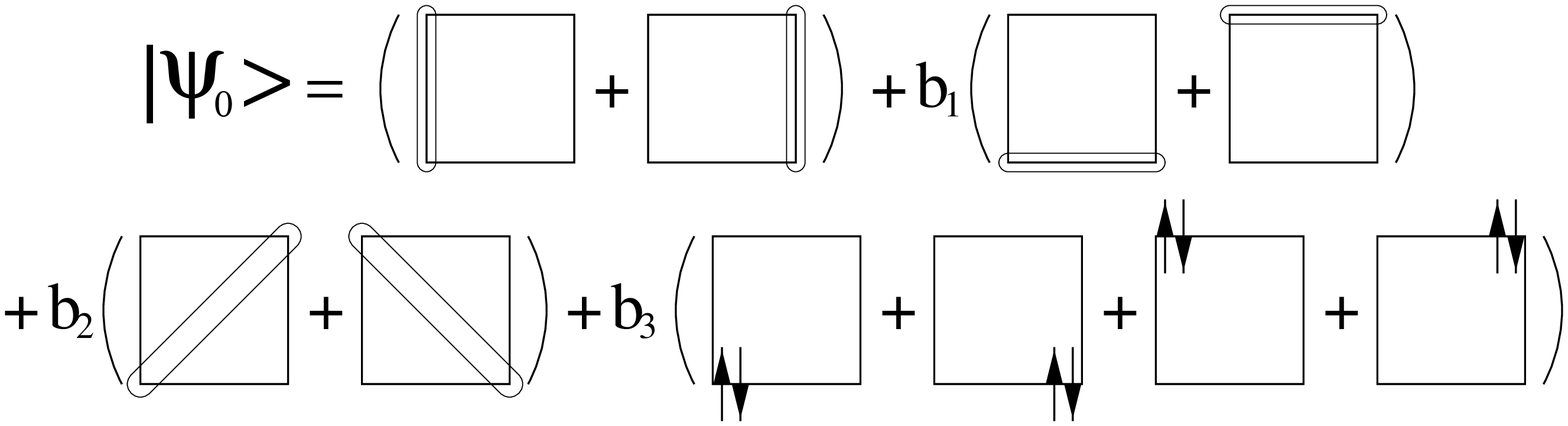}}
\caption{Ground state wave function for the $2 \times 2$ plaquette with 
two holes.}
\label{fig:2x2holes}
\end{figure}
\vspace{.25in}

\noindent
background captures the
essential physics of the $t-J$ ladder at small dopings.  Since the 
$t-J$ model is the large U limit of the Hubbard model, we expect this
picture to hold for the Hubbard model at large U and small dopings.
Therefore, we consider an Ansatz of hole pairs moving through our 
generalized RVB background for the Hubbard ladder at small dopings.

Since the structure of the hole pairs is based on the exact solution
to the 2x2 plaquette with 2 holes, we consider the 2x2 case in detail
below.

\subsection{The 2x2 Plaquete with Two Holes}
The ground state of the 2x2 plaquette with 2 holes is given by
\begin{equation}
 \mid \psi_0^h \rangle \  =  \ \mid \varphi_0^h \rangle  +  
                     b_1~\mid \varphi_1^h \rangle  + 
		     b_2~\mid \varphi_2^h \rangle +
		     b_3~\mid \varphi_3^h \rangle  
\end{equation}
where
\begin{eqnarray}
 \mid \varphi_0^h \rangle & = &
     \Delta^{\dagger}_{(1,1),(1,2)} \mid 0 \rangle + 
     \Delta^{\dagger}_{(2,1),(2,2)} \mid 0 \rangle \, ,
 \nonumber \\
 \mid \varphi_1^h \rangle & = &
     \Delta^{\dagger}_{(1,1),(2,1)} \mid 0 \rangle + 
     \Delta^{\dagger}_{(1,2),(2,2)} \mid 0 \rangle  \, ,
 \nonumber \\
 \mid \varphi_2^h \rangle & = & 
     \Delta^{\dagger}_{(1,1),(2,2)} \mid 0 \rangle + 
     \Delta^{\dagger}_{(2,1),(1,2)} \mid 0 \rangle \, ,
 \nonumber \\
 \mid \varphi_3^h \rangle & = &
     D^{\dagger}_{(1,1)} \mid 0 \rangle  + 
     D^{\dagger}_{(1,2)} \mid 0 \rangle 
 \nonumber \\
 & + &  D^{\dagger}_{(2,1)} \mid 0 \rangle  + 
     D^{\dagger}_{(2,2)} \mid 0 \rangle   \, . 
\end{eqnarray}
$\mid \psi_0^h \rangle$ is shown schematically in Fig.~\ref{fig:2x2holes}.

In Table \ref{table:2x2holes} we list the values of the parameters 
for several values of $U$ (with $t=t_{\perp} = 1$).

Notice that the solution to the 2x2 case (with $t=t_{\perp}$)
has $D_4$ symmetry.
However, now the ground state transforms in the {\it scalar}
representation of $D_4$.  Recalling     \\

\begin{table}
\caption{Values of the parameters for the $2 \times 2$ plaquette 
with two holes (with $t = t_{\perp} = 1$) which gives the exact 
ground state.} 
\begin{center}
\begin{tabular}{|c|c|c|c|c|c|c|c|c|} 
\hline 
$U$ &  $b_1$ & $b_2$ & $b_3$ \\ \hline \hline

 $8$  & $1.0$ & $1.2470$ & $0.3569$  \\ \hline 

 $16$ & $1.0$ & $1.3131$ & $0.2100$   \\ \hline 

 $24$ & $1.0$ & $1.3420$ & $0.1483$  \\ \hline 
\end{tabular}
\end{center}
\label{table:2x2holes}
\end{table}
\vspace{.2in}

\noindent
that the ground state of the half-filled ladder transforms  
in the $B_2$ representation, we
see that the operator that creates a hole pair out of the
undoped system has $d_{x^2-y^2}$ symmetry. \cite{trugman}  These
facts continue to be true for low doping in larger ladders. 
Also, notice that
$b_3 = {\cal O}(\alpha)$ where $\alpha$ is from the half-filled 
ladder.  This will play a role in writing our Ansatz for the 
doped ladder.
 
 
\subsection{The Ladder}
\begin{figure}
\epsfxsize= 2.9in
\epsfysize= .6in
\centerline{\epsfbox{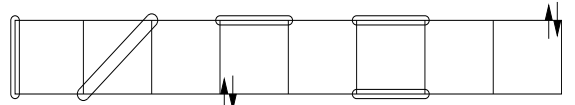}}
\caption{Typical configuration appearing in the ground state 
wave function of the doped ladder.}
\label{fig:typicaldoped}
\end{figure}

Using the configurations of the generalized RVB state as well as
the hole pair configurations, a typical configuration for the 
doped ladder is shown in Fig.~\ref{fig:typicaldoped}.

\vspace{.15in}

\vspace{.2in}
\noindent
The ground state will be a superposition of all such configurations 
shown in Fig.~\ref{fig:typicaldoped}.  Fortunately, we can generate 
such a state {\it recursively}.  
Specifically, we build the ground state of a ladder with 
$N+\nu$ rungs and $P+\mu$ holes using the knowledge of the ground
states of a ladder with $N, N+1, \ldots , N+\nu  -1$ rungs and
$P, P+1, \ldots, P+\mu$ holes.  This is achieved by recursion relations
which express the ground state $\mid N+\nu, P+\mu \rangle$ in terms
of the ground states 
$\{\mid N+i, P+j \rangle \}$ with $i=0, \ldots, \nu-1$ and 
$j=0, \ldots, \mu$.\cite{sierra2}

Using the above ingredients, we consider the following Ansatz
for the doped Hubbard ladder
\breakon
\begin{eqnarray}
 \mid N+2~P+1 \rangle\ &  = & \ 
      \mid \phi_0 \rangle_{N+2} \mid N+1~P+1 \rangle
   \, + \, \alpha \mid \phi_1 \rangle_{N+2} \mid N+1~P+1 \rangle  
   \, + \, \beta \mid \phi_2 \rangle_{N+1,N+2} \mid N~P+1 \rangle
 \nonumber \\
  & + & \, \gamma \mid \phi_3 \rangle_{N+1,N+2} \mid N~P+1 \rangle
   \, + \, \xi \mid \phi_4 \rangle_{N+1,N+2} \mid N~P+1 \rangle 
   \, + \, \eta \mid \phi_5 \rangle_{N+1,N+2} \mid N~P+1 \rangle
 \nonumber \\
  & + & \, \delta \mid \phi_6 \rangle_{N,N+1,N+2} \mid N-1~P+1 \rangle
   \, + \, \varepsilon \mid \phi_7 \rangle_{N,N+1,N+2} \mid N-1~P+1 \rangle
 \nonumber \\
  & + & \,  \mid \phi_0^h \rangle_{N+2} \mid N+1~P \rangle
   \, + \, \lambda \mid \phi_1^h \rangle_{N+1,N+2} \mid N~P \rangle
   \, + \, \zeta \mid \phi_2^h \rangle_{N+1,N+2} \mid N~P \rangle
 \nonumber \\
  & + & \, \mu \mid \phi_3^h \rangle_{N,N+1,N+2} \mid N-1~P \rangle
   \, + \, \nu \mid \phi_4^h \rangle_{N,N+1,N+2} \mid N-1~P \rangle 
\label{dopedansatz}
\end{eqnarray}
\breakoff

\vspace{.2in}
\begin{figure}
\epsfxsize= 3.6in
\epsfysize= 2.9in
\centerline{\epsfbox{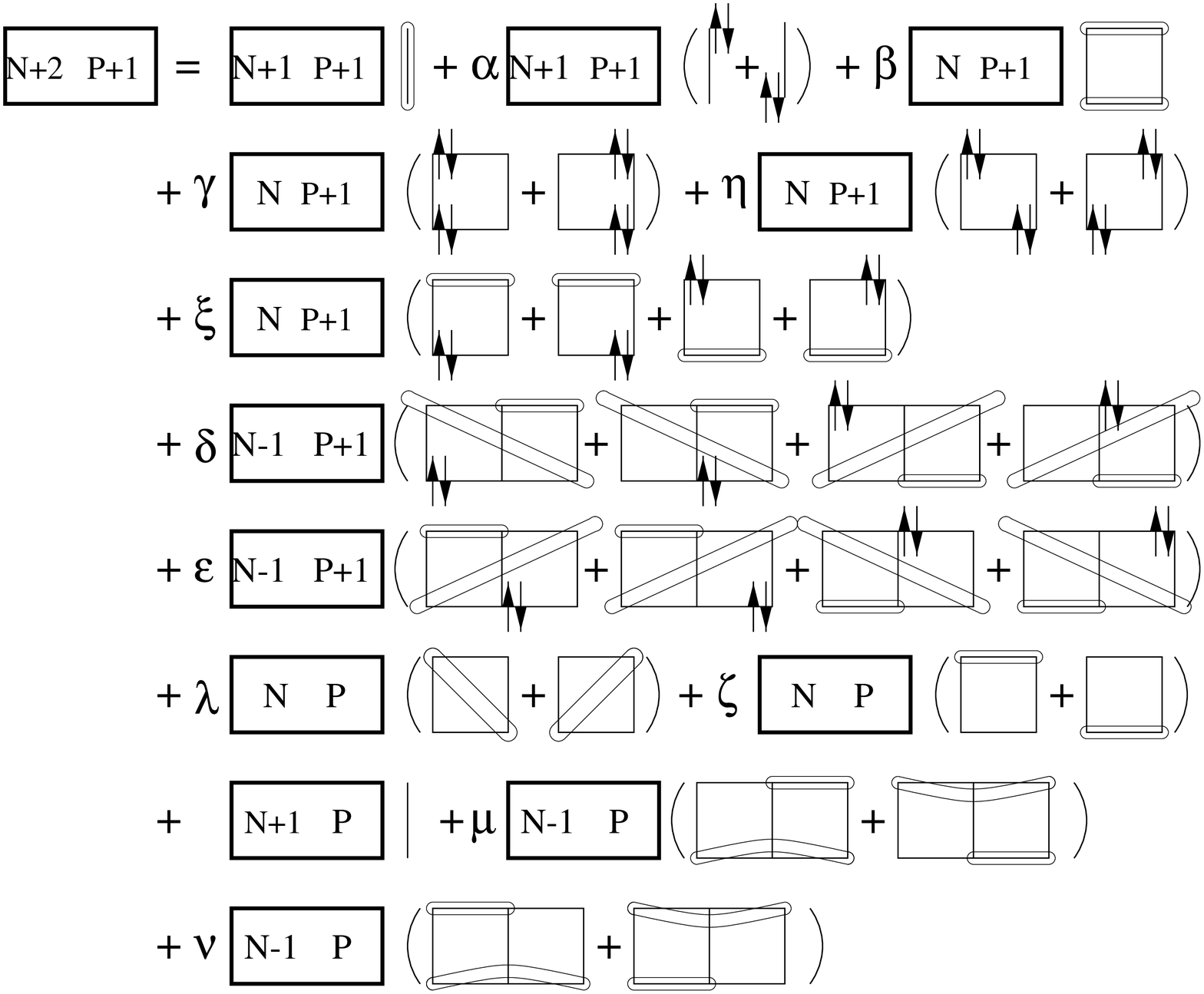}}
\caption{The RVA Ansatz for the doped Hubbard ladder.}
\label{fig:goodstates1}
\end{figure}
\vspace{.25in}

\noindent
which is shown schematically in Fig.~\ref{fig:goodstates1}.
For completeness, the states in terms of the operators in
Eqs.~\ref{singlet} and \ref{double} are given in Appendix B.

A few words are in order about the configurations in our Ansatz.
i) The Ansatz contains the configurations for the RVB state, as well
as the configurations for holes bound in pairs.
The hole pair states are based on the exact solution to the 2x2
plaquette with 2 holes, and our ansatz reproduces this exact solution.
ii) The states 
$\mid \phi_3^h \rangle_{N,N+1,N+2}$ and 
$\mid \phi_4^h \rangle_{N,N+1,N+2}$ extend over three rungs. 
These states are necessary to allow the hole pairs 
to move smoothly through the RVB background.  
iii) Note that physically, the picture of hole pairs moving through 
an RVB background can only be appropriate for low dopings.

To compute the values of the coefficients in our Ansatz, we 
treat them as variational parameters and minimize the ground-state 
energy with respect to these parameters.  It will be useful to define 
\begin{eqnarray}
 E_{N,P} \ &  = & \ \langle N~P \mid H_N \mid N~P \rangle \, ,  \nonumber \\
 D_{N,P} \ &  = & \ \langle N-1~P \mid_N \langle \phi_0 \mid 
                 H_N \mid N~P \rangle \, ,  \nonumber \\
 C_{N,P} \ &  = & \ \langle N-1~P \mid_N \langle \phi_1 \mid 
                 H_N \mid N~P \rangle \, ,  \nonumber \\
 Z_{N,P} \ &  = & \ \langle N~P \mid N~P \rangle \, ,  \nonumber \\
 Y_{N,P} \ &  = & \ \langle N-1~P \mid_N \langle \phi_0 \mid 
                 N~P \rangle \, ,  \nonumber \\
 X_{N,P} \ &  = & \ \langle N-1~P \mid_N \langle \phi_1 \mid 
                 N~P \rangle \, .  
\end{eqnarray}
They are supplemented by the initial conditions
\begin{eqnarray}
& &  Z_{N,P=N}=1, Y_{N,P=N}=0, X_{N,P=N}=0, \nonumber \\
& &  E_{N,P=N}=0, D_{N,P=N}=0, C_{N,P=N}=0,  \nonumber \\
& &  F_{N<P,P}=0 \ {\rm for} \ F = Z, Y, X, E, D, C \, . 
\end{eqnarray}

To determine the values for the variational parameters for given
(finite) values of $N$ and $P$, we iterate the recursion relations
and we minimize the quantity $E_{N,P} / Z_{N,P}$.
The actual recursion relations are quite unwieldly; we have relegated 
them, as well as their derivation, to Appendix B.

What is the nature of the state we have constructed?  
In order to answer this question, we plot $\beta$, $\lambda$, 
and $\zeta$ vs. doping.
These parameters contain most of the physics of our Ansatz. 
$\beta$ is the ``RVB parameter''; $\lambda$ and $\zeta$ are
the weights of the hole pair configurations.  The results were
obtained on a $2 \times 32$ ladder.

\begin{figure}
\epsfxsize= 3.25in
\centerline{\epsfbox{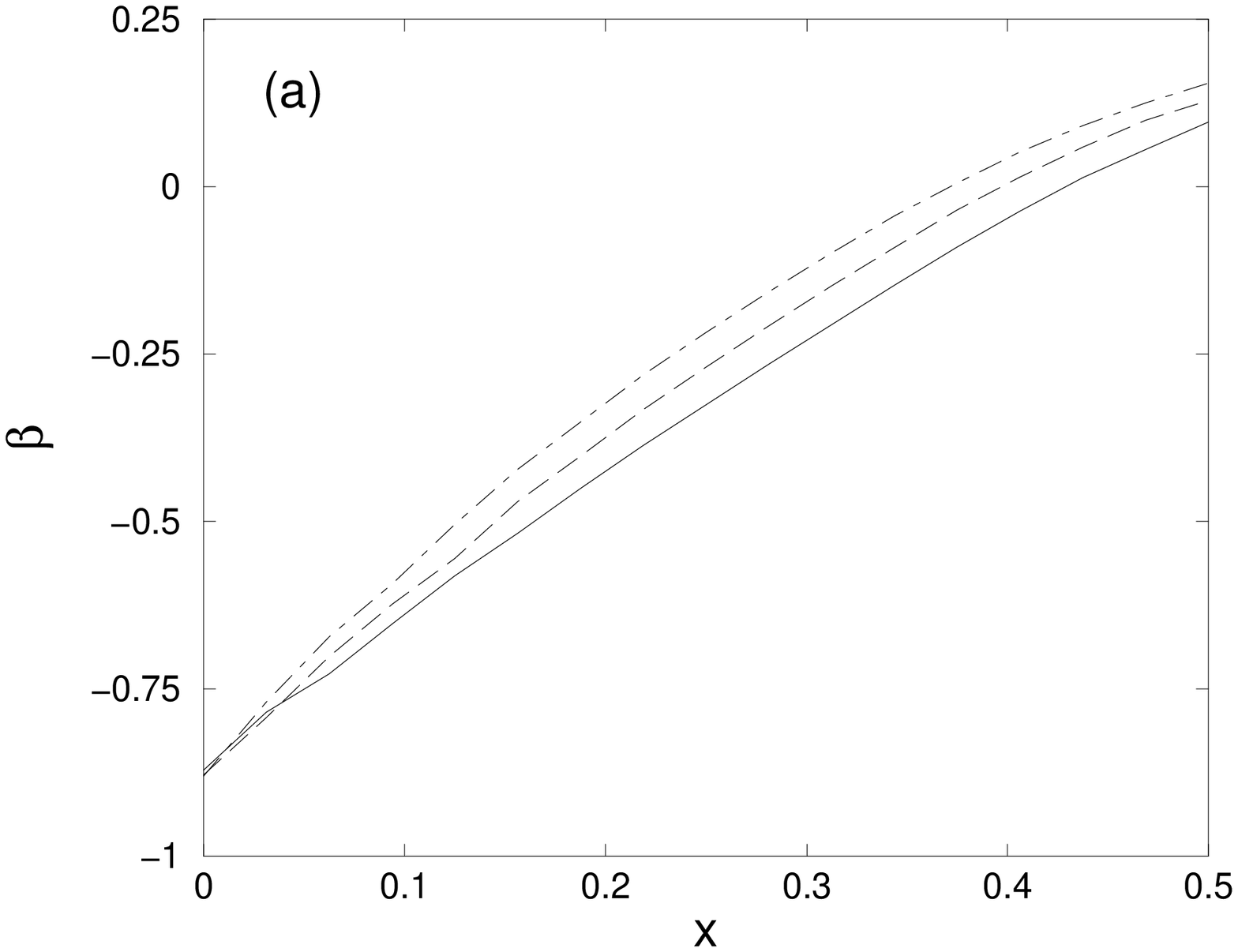}}
\epsfxsize= 3.25in
\centerline{\epsfbox{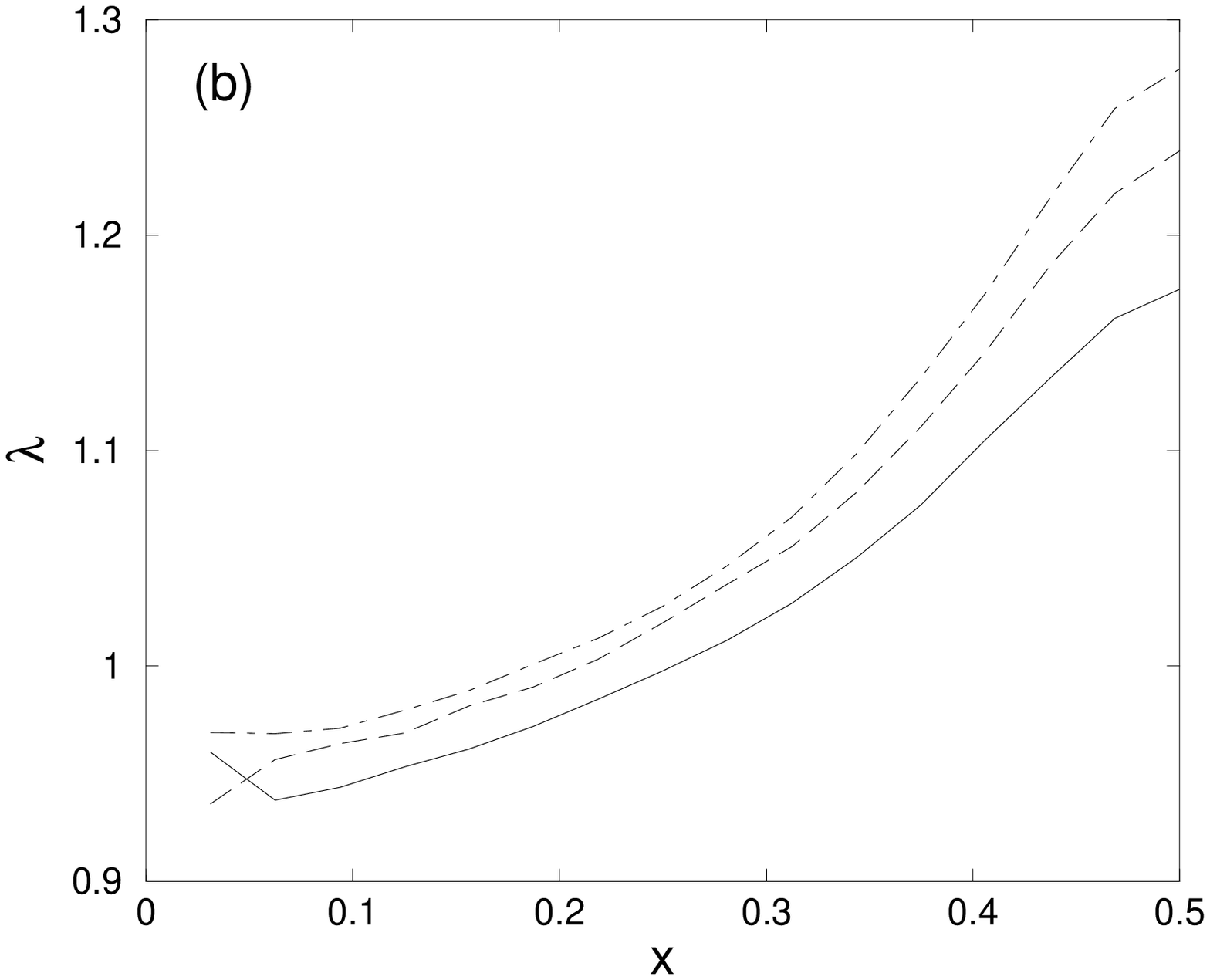}}
\epsfxsize= 3.25in
\centerline{\epsfbox{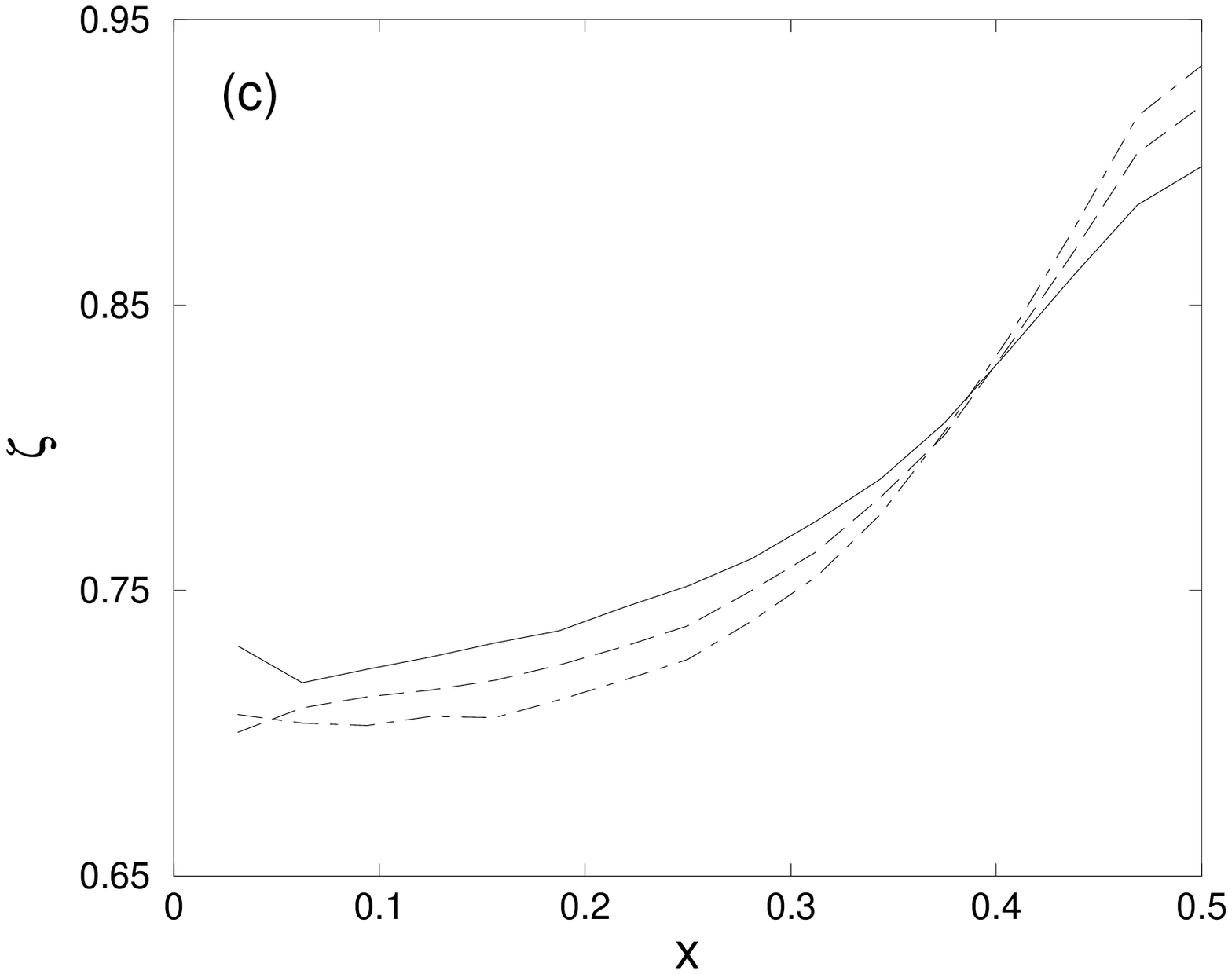}}
\caption{(a) $\beta$ vs. doping, $x$, for a $2 \times 32$ ladder at
U=8 (solid line), U=12 (dashed line), and U=16 (dashed-dotted line.
(b) Same as (a) for $\lambda$.  (c) Same as (a) for $\zeta$.}
\label{fig:parameters}
\end{figure}

First, consider Fig.~\ref{fig:parameters}(a).
$\beta$ begins at ${\cal O}(-1)$ and increases (i.e., becomes less
negative) with doping until a critical doping, $x_c$, where it vanishes.  
Beyond this doping, $\beta$ is positive.  This has also been found for
the $t-J$ ladder. \cite{sierra2} Upon doping, the hole
pairs cause destructive interference which degrades the RVB 
mechanism.  For $x > x_c$, this destructive interference has
driven $\beta$ positive, and it is no longer appropriate to think of
our state as describing hole pairs moving through an RVB background.
\cite{sierra2}  

Similar to the t-J ladder, the difference between 
$x < x_c$ and $x > x_c$ can be attributed to 
two different internal structures of the hole pairs.  For $x < x_c$
the hole pairs have a $d_{x^2-y^2}$ structure {\it relative} to the 
RVB background.  For $x > x_c$ the hole pairs have an $s$-wave like
symmetry relative to their background. \cite{sierra2}

Now consider Figs.~\ref{fig:parameters}(b) and \ref{fig:parameters}(c).  
First of all, notice that $\lambda > \zeta$.  This shows the importance 
of the diagonal frustrating bonds for all dopings.  \cite{diagonal}
Also, notice that 
$\lambda$ and $\zeta$ both reach their maximum at $x=1/2$.  At $x=1/2$
the system is essentially a large scale reproduction of the 2x2
plaquette with 2 holes. \cite{sierra2}  Indeed, the values of 
$\lambda$ and $\zeta$ at $x=1/2$ are similar to their values
for the 2x2 plaquette.

\vspace{.2in}


\section{Ground State Energies}

\begin{figure}
\epsfxsize= 3.25in
\centerline{\epsfbox{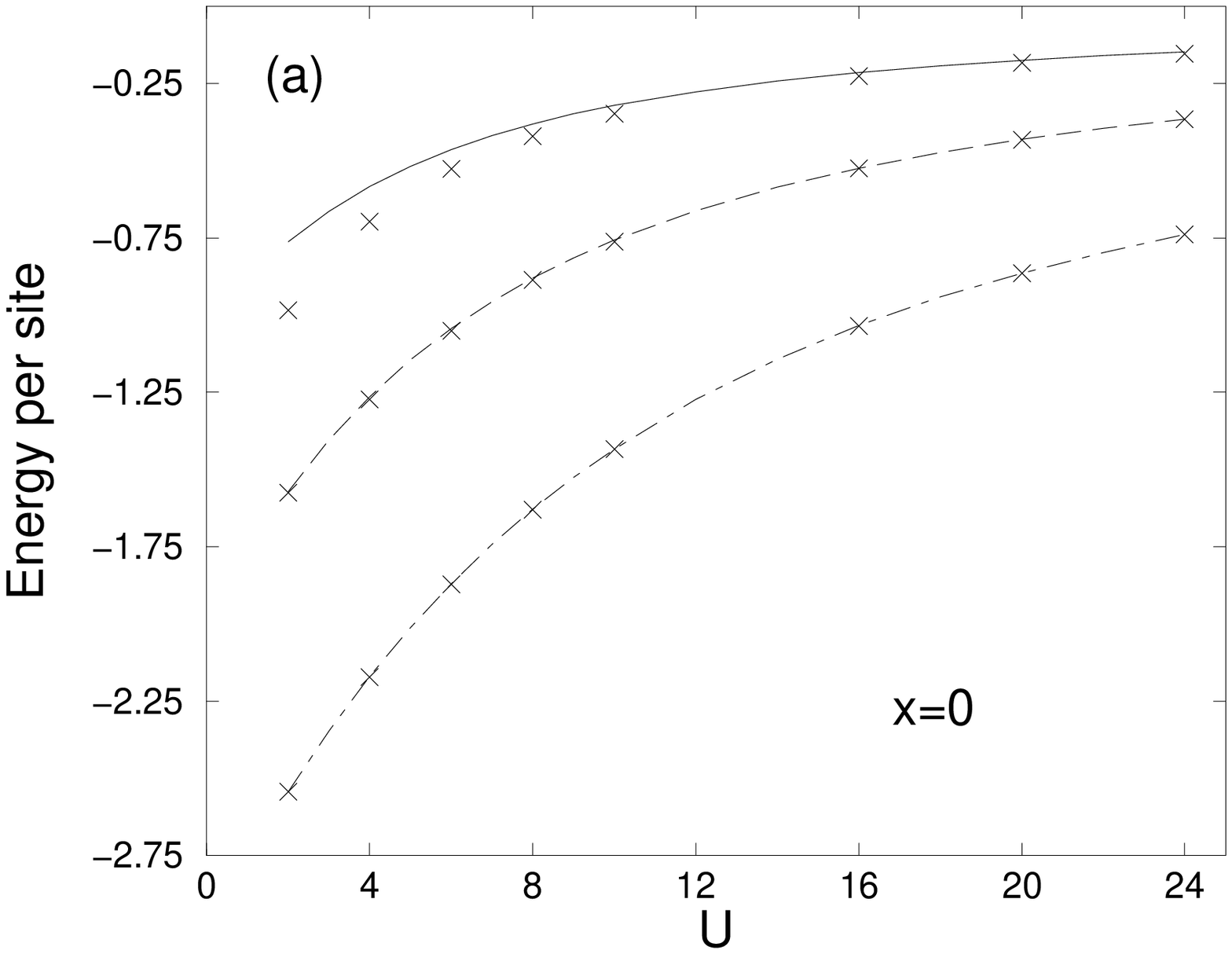}}
\epsfxsize= 3.25in
\centerline{\epsfbox{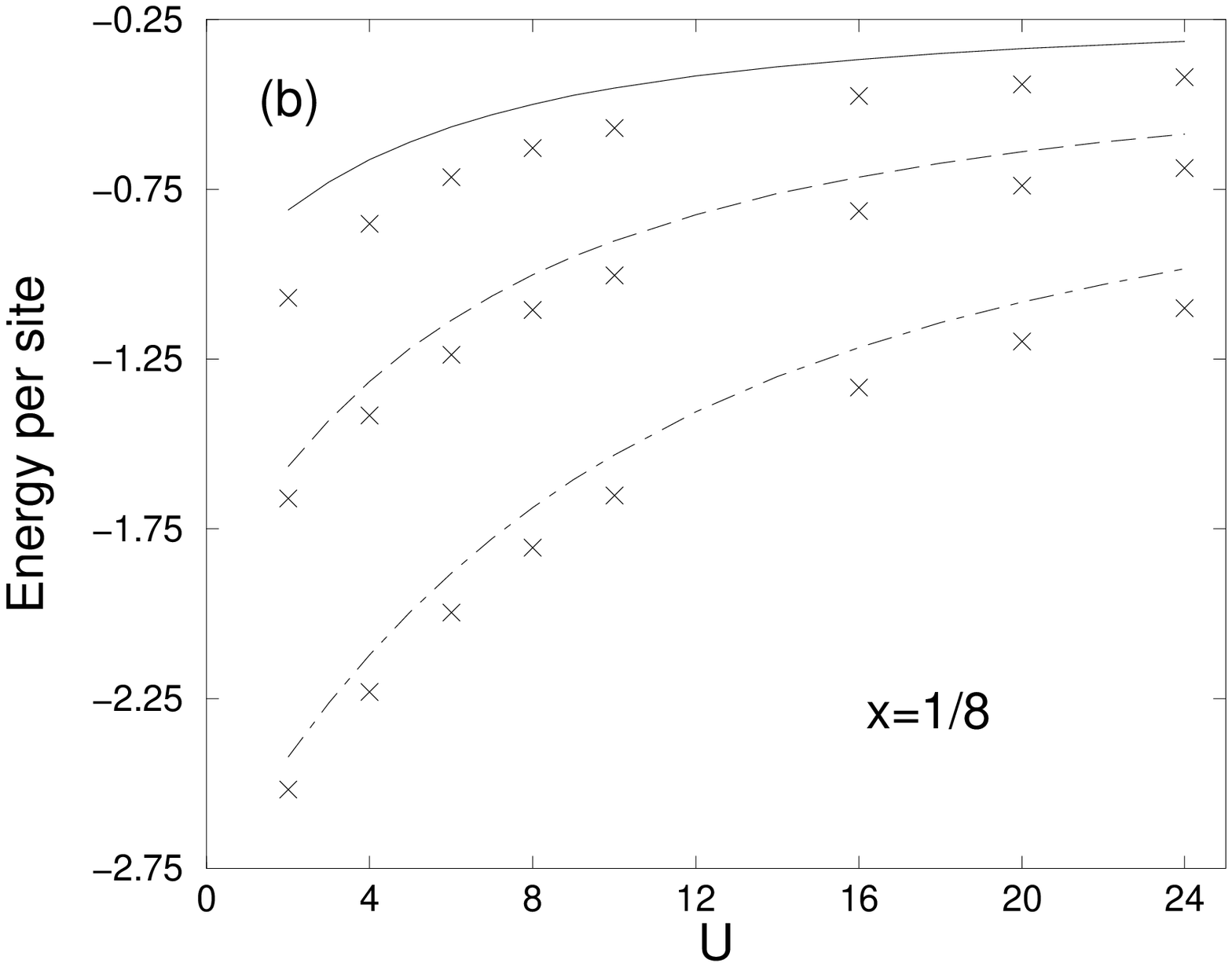}}
\epsfxsize= 3.25in
\centerline{\epsfbox{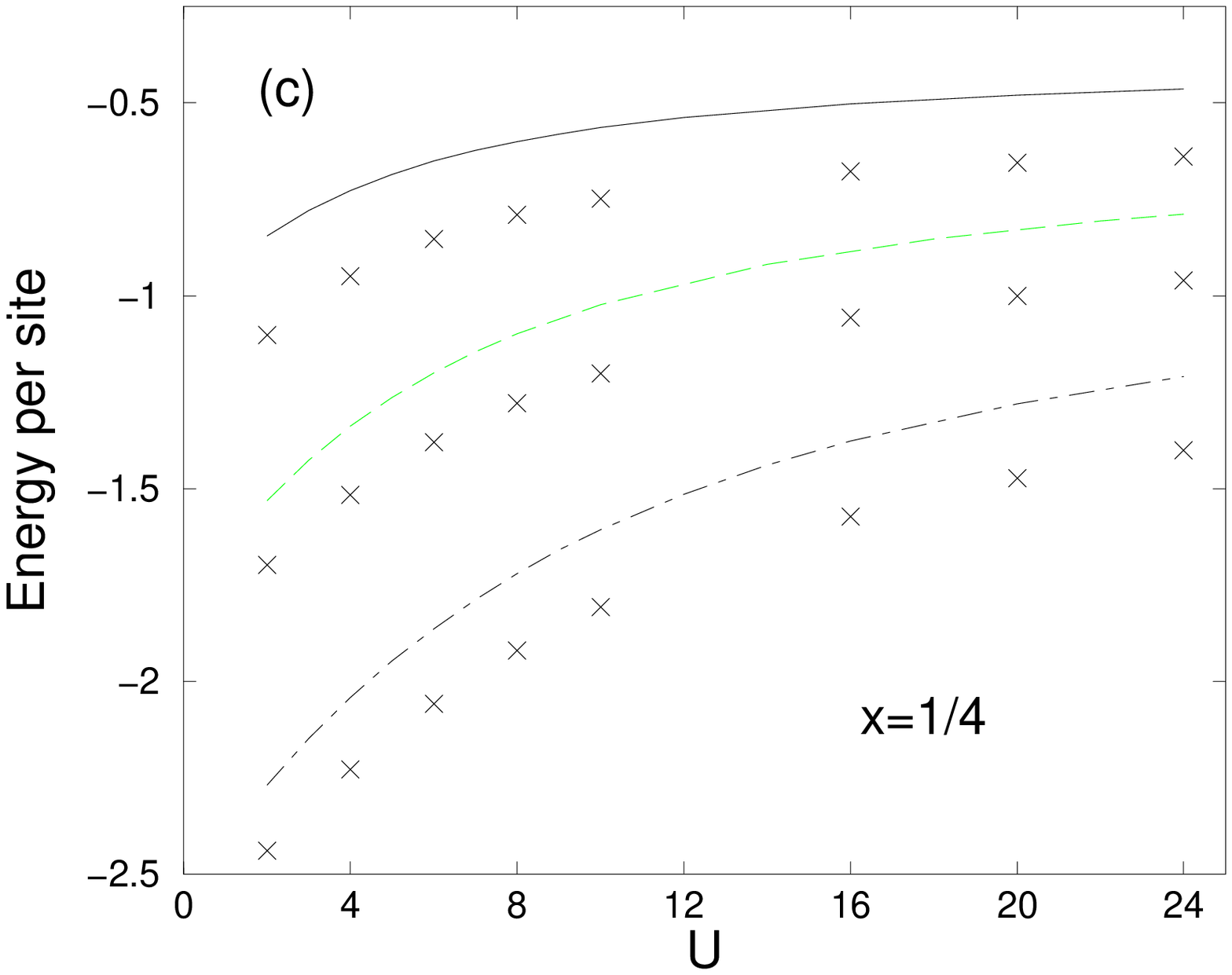}}
\caption{(a) Ground state energy per site at half-filling ($x=0$)
vs. $U$ for $t_{\perp} = 1$ (solid line), $t_{\perp} = 2$ (dashed line), 
$t_{\perp} = 3$ (dashed-dotted line).  For comparison, DMRG results 
(shown as $\times$) are presented for the same set of parameters.  
(b) Same as (a) except for $x=1/8$.  (c) Same as (a) except for $x=1/4$.}
\label{fig:energyvsU}
\end{figure}

First, we show results for Energy per site vs. $U$ in 
Fig.~\ref{fig:energyvsU}.  For comparison, DMRG results are presented
for the same set of parameters. \cite{reinhardt}
At half-filling, as we would expect, our Ansatz is most accurate for 
large $U$ and large $t_{\perp}$.  The ground state energy per site
for a $2 \times 32$ half-filled ladder as a function of $U$ for
various $t_{\perp}$ is shown in Fig.~\ref{fig:energyvsU}(a).  
For $U=8$ and $t_{\perp}=1$, the energy from the
RVA agrees with DMRG to within $90 \%$ and improves as $U$ or
$t_{\perp}$ is increased.  Up to about $U=10$, longer bonds
(extending over at least 3 rungs) are coming into play.  These
states should be included in the Ansatz to further improve the 
overlap with the ground state.  At $U=16$ and $t_{\perp}=1$, our ansatz 
gives a ground state energy within $94 \%$ of the DMRG result.   

It should be noted that for the Heisenberg ladder, the   
RVB state gives a ground state energy within $94 \%$ of true  \\

\vspace{.25in}

\noindent
ground state energy, obtained from DMRG. \cite{sierra1}  A recent \\

\begin{figure}
\epsfxsize= 3.0in
\centerline{\epsfbox{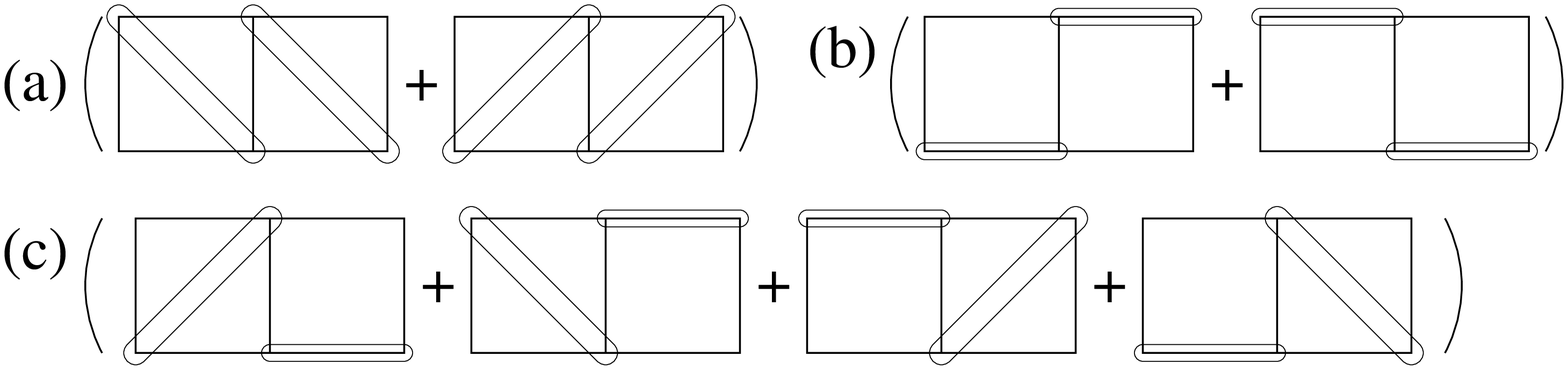}}
\caption{``Pair-breaking'' configurations which are playing a rather
large role in the ground state of the doped Hubbard ladder.}
\label{fig:pairbreaking}
\end{figure}
\vspace{.25in}

\noindent
DMRG study of different 
ladder models found that the Hubbard model and 
Heisenberg model begin to agree only for rather large $U$ ($U \approx 16$).  
\cite{jeckelmann}  Therefore, it is not surprising that the RVB picture 
becomes as good for the Hubbard model as it is for the Heisenberg model at
$U \approx 16$.

Fig.~\ref{fig:energyvsU}(b) shows the ground state energy of the 
RVA Ansatz as a funtion of $U$ for various $t_{\perp}$ for a doping
of $x=1/8$ on a $2 \times 32$ ladder.  Again, we show energies obtained
from DMRG for the same set of parameters.  For $U=16$ and $t_{\perp} = 1$,
the two energies agree to only within $77 \%$ and improves slightly as
$t_{\perp}$ is increased.  For example, at $U=16$ and $t_{\perp} = 2$,
the overlap of energies increases to $87 \%$.  Further discrepancies
occur when the doping is increased to $x=1/4$ (see 
Fig.~\ref{fig:energyvsU}(c) ).

\begin{figure}
\epsfxsize= 2.9in
\centerline{\epsfbox{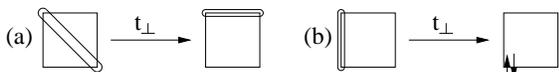}}
\caption{For $t << t_{\perp}$, hopping along the rung dominates.  By 
putting the particles (or holes) on adjacent rungs, we can have the
situation shown in (a).  However, by putting both particles (or holes)
on the same rung, we get the situation shown in (b), which is 
energetically unfavorable.}
\label{fig:adjacent}
\end{figure}

The differences in ground state energies occur due to the importance
of ``pair-breaking'' configurations, like those shown in
Fig.~\ref{fig:pairbreaking}.  The weights of these types of states
increase as we move away from half-filling.  Consequently, to
further improve the RVA Ansatz, such states must be included in the
wave function.  Note also that our Ansatz would be essentially exact 
for the case where hole pairs are well localized on a rung.  For the
$t-J$ model with $J_{rung} \gg J_{chain}, t$, pairs are well localized
along a rung, and the ground state is essentially a product of 
rung singlets and rung hole pairs.  
However, for the Hubbard model at strong coupling 
(i.e., $U \gg t_{\perp}, t$), this is not the case.
Holes would always rather occupy {\it adjacent} rungs, even 
for $t_{\perp} >> t$, since this minimizes the Coulomb energy from 
doubly occupied sites.  To see this consider a $2 \times 2$ plaquette
with 2 holes; let $t \ll t_{\perp}$ and $t, t_{\perp} \ll U$.  
With 1 particle on each rung (i.e., one hole on each rung), the 
ground state energy is approximately $-2 t_{\perp}$; with both particles 
on the same rung (i.e., both holes on the same rung), the ground state
energy is approximately $-J = -4t_{\perp}^2 / U$.  Therefore, 
at large U, the particles would rather occupy adjacent rungs.
(See Fig.~\ref{fig:adjacent}.)

The situation we have with the doped ladder is similar to what we
had for the half-filled ladder in the early stages of this work.
We found that without the states \\

\vspace{.2in}
\vspace{.25in}

\begin{figure}
\epsfxsize= 3.25in
\centerline{\epsfbox{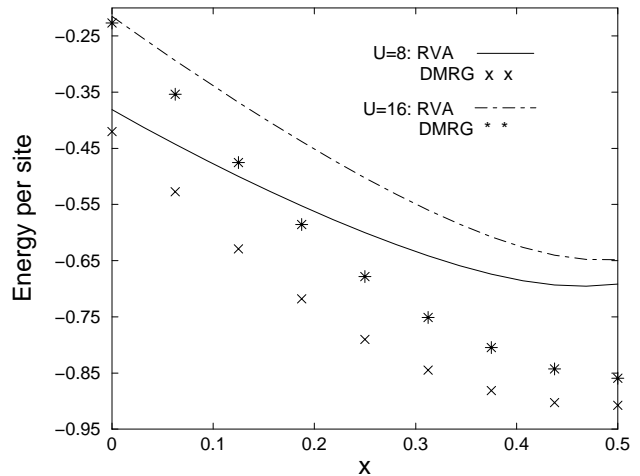}}
\caption{Ground state energy per site vs. doping for a $2 \times 32$ ladder
(with $t = t_{\perp} = 1$)for $U=8$ (solid line) and $U=16$ 
(dashed-dotted line).  For comparison, DMRG results for $U=8$ ($\times$)
and $U=16$ (*) are shown.}
\label{fig:xeplot1}
\end{figure}
\vspace{.25in}

\noindent
 $\mid \phi_6 \rangle$ and 
 $\mid \phi_7 \rangle$ which extend over three rungs 
(see Fig.~\ref{fig:goodstates}), the RVA did not accurately reproduce
the ground state even at extremely large $U$.  However, once we included 
$\mid \phi_6 \rangle$ and  $\mid \phi_7 \rangle$, the results from the 
RVA improved drastically.  Based on these results, we expect the RVA 
to greatly improve by including the states shown in 
Fig.~\ref{fig:pairbreaking}.   

In Fig.~\ref{fig:xeplot1} we plot Energy per site vs. Doping 
for a 
$2 \times 32$ ladder for $U=8$ and $U=16$ (with $t_{\perp} = 1$)
in order to better understand the region of validity of our RVA ansatz.
Again, we see good agreement with DMRG results at half-filling.  However,
as soon as we dope, configurations like those shown in 
Fig.~\ref{fig:pairbreaking} are also important.

\vspace{.2in}
It is interesting to note that the idea of hole pairs moving through
the RVB background seems to more accurately represent the ground state
of the $t-J$ model than the Hubbard model.  Using the 
well known relation at strong coupling, 
$J \approx \frac{4t^2}{U}$, the RVA agrees to within $92 \%$ of the 
true ground state energy of the $t-J$ ladder for $J=0.5$ $(U=8)$ at a
doping of $x=1/8$.  
There are two ways to interpret this; either the t-J model supports pairing 
better than the Hubbard model, or we must view the hole pairs in the Hubbard 
model as having a larger size (i.e. larger coherence length.)


\section{Concluding Remarks}

To summarize, we applied the Recurrent Variational Approach to the 
two-leg Hubbard ladder.  Our results were in qualitative agreement 
with previous results on the Heisenberg and t-J ladders.  For the 
half-filled ladder, the generalized RVB state became more accurate
in the parameter regime where the Hubbard and Heisenberg ladders
were shown to coincide.  However, 
comparison of the RVA with DMRG for the doped ladder indicate that 
hole-pairs moving through an RVB background is incomplete; it does not
capture the essential physics.  ``Pair-breaking'' configurations are 
also necessary to capture the essential physics.

As we saw, the strength of the RVA is the ease in which we could 
extract the physics.  We were able to see the importance of the 
configurations in our Ansatz quite easily.  Furthermore, the RVA 
has a natural way in which to include longer bonds in the Ansatz
to more accurately represent the groundstate wavefunction.
The importance of such additional states to the physics of the ladder
is not easily probed with other techniques.

Generalized RVB states similar to ours have been
considered previously for the half-filled Hubbard ladder.
\cite{previous}  Fano {\it et. al.} were even able to produce
an Ansatz coming within $98 \%$ of the true groundstate energy for 
a (half-filled) 2x4 ladder at $U=16$. 
(Their Ansatz included diagonal bonds of length $\sqrt{5}$.)  However,
none of these works considered the doped case.  Using the approach
in Ref. 17 (in terms of dimer coverings), it appears to be a formidable 
task to consider doping.  This is one of the strengths of the RVA; doping
is handled rather easily.  Even though our results for the doped ladder 
showed that hole pairs moving through an RVB background is incomplete,
the RVA offers a straightforward way to improve the situation,
namely include ``pair-breaking'' configurations (shown in 
Fig.~\ref{fig:pairbreaking} ) in the Ansatz.

Another (and probably better) way to improve the situation for 
the doped ladder is to consider a Matrix Product Ansatz. 
\cite{matrix1,matrix2}  A Matrix Product Ansatz can be generated by 
first order recursion relations. \cite{matrix2} In the RVA, the size 
of the hole-pairs are fixed.  (In our case, the hole pairs had a size of 
one lattice spacing.)  However, by construction, the Matrix Product Ansatz 
takes into account hole pairs of {\it arbitrary} size.  We leave this 
(and other possibilities) for future work.


\vspace{.25in}
\section*{Acknowledgments}
We would like to thank D.~J. Scalapino, S. Daul, S. Hortikar, R. Konik, 
C.~L. Martin, and M.~A. Martin-Delgado for helpful discussions 
and comments.  EHK gratefully acknowledges the warm hospitality 
of Argonne National Laboratory where parts of this manuscript were 
written.  DD would like to thank S. Daul and R.~M. Noack for
assistance with the DMRG results presented here.  EHK was supported by 
NSF grant No.~DMR-9527304, GS by DGES grant PB97-1190, and DD by DOE 
grant No. DE-FG03-85ER451907.


\appendix
\section{The Half-Filled Ladder}

The states in the Ansatz for the half-filled ladder of 
Eq.~\ref{halffilledansatz}  are given by
\breakon
\begin{eqnarray}
 \mid \phi_0 \rangle_{N+2} & = &
     \Delta^{\dagger}_{(N+2,1),(N+2,2)} \mid 0 \rangle_{N+2} \, ,
 \nonumber \\
 \mid \phi_1 \rangle_{N+2} & = & 
     D^{\dagger}_{(N+2,1)} \mid 0 \rangle_{N+2} +
     D^{\dagger}_{(N+2,2)} \mid 0 \rangle_{N+2} \, ,
 \nonumber \\
 \mid \phi_2 \rangle_{N+1,N+2} & = &
     \Delta^{\dagger}_{(N+1,1),(N+2,1)} 
     \Delta^{\dagger}_{(N+1,2),(N+2,2)} \mid 0 \rangle_{N+1,N+2} \, ,
 \nonumber \\
 \mid \phi_3 \rangle_{N+1,N+2} & = &
     D^{\dagger}_{(N+1,1)} D^{\dagger}_{(N+1,2)} \mid 0 \rangle_{N+1,N+2} +  
     D^{\dagger}_{(N+2,1)} D^{\dagger}_{(N+2,2)} \mid 0 \rangle_{N+1,N_2} \, ,
 \nonumber \\
 \mid \phi_4 \rangle_{N+1,N+2} & = &
     \Delta^{\dagger}_{(N+1,1),(N+2,1)} D^{\dagger}_{(N+1,2)} 
     \mid 0 \rangle_{N+1,N+2} +
     \Delta^{\dagger}_{(N+1,1),(N+2,1)} D^{\dagger}_{(N+2,2)} 
     \mid 0 \rangle_{N+1,N+2}  
 \nonumber \\
& + &  \Delta^{\dagger}_{(N+1,2),(N+2,2)} D^{\dagger}_{(N+1,1)} 
     \mid 0 \rangle_{N+1,N+2} +
     \Delta^{\dagger}_{(N+1,2),(N+2,2)} D^{\dagger}_{(N+2,1)} 
     \mid 0 \rangle_{N+1,N+2} \, ,
 \nonumber \\
 \mid \phi_5 \rangle_{N+1,N+2} & = &
     D^{\dagger}_{(N+1,1)} D^{\dagger}_{(N+2,2)} 
     \mid 0 \rangle_{N+1,N+2} +
     D^{\dagger}_{(N+1,2)} D^{\dagger}_{(N+2,1)} 
     \mid 0 \rangle_{N+1,N+2}  \, ,
 \nonumber \\
 \mid \phi_6 \rangle_{N,N+1,N+2} & = &
  \Delta^{\dagger}_{(N,1),(N+2,2)} \Delta^{\dagger}_{(N+1,1),(N+2,1)}
  D^{\dagger}_{(N,2)} \mid 0 \rangle_{N,N+1,N+2}  \nonumber \\
 & + & \Delta^{\dagger}_{(N,1),(N+2,2)} \Delta^{\dagger}_{(N+1,1),(N+2,1)}
  D^{\dagger}_{(N+1,2)} \mid 0 \rangle_{N,N+1,N+2}  \nonumber \\ 
 & + & \Delta^{\dagger}_{(N,2),(N+2,1)} \Delta^{\dagger}_{(N+1,2),(N+2,2)}
  D^{\dagger}_{(N,1)} \mid 0 \rangle_{N,N+1,N+2} \nonumber \\
 & + & \Delta^{\dagger}_{(N,2),(N+2,1)} \Delta^{\dagger}_{(N+1,2),(N+2,2)}
  D^{\dagger}_{(N+1,1)} \mid 0 \rangle_{N,N+1,N+2} \, ,  
 \nonumber \\
 \mid \phi_7 \rangle_{N,N+1,N+2} & = &
  \Delta^{\dagger}_{(N,1),(N+2,2)} \Delta^{\dagger}_{(N,2),(N+1,2)}
  D^{\dagger}_{(N+1,1)} \mid 0 \rangle_{N,N+1,N+2}  \nonumber \\
 & + & \Delta^{\dagger}_{(N,1),(N+2,2)} \Delta^{\dagger}_{(N,2),(N+1,2)}
  D^{\dagger}_{(N+2,1)} \mid 0 \rangle_{N,N+1,N+2}  \nonumber \\ 
 & + & \Delta^{\dagger}_{(N,2),(N+2,1)} \Delta^{\dagger}_{(N,1),(N+1,1)}
  D^{\dagger}_{(N+1,2)} \mid 0 \rangle_{N,N+1,N+2}  \nonumber \\
 & + & \Delta^{\dagger}_{(N,2),(N+2,1)} \Delta^{\dagger}_{(N,1),(N+1,1)}
  D^{\dagger}_{(N+2,2)} \mid 0 \rangle_{N,N+1,N+2} \, .
\end{eqnarray}

To derive the recursion relations, the following inner products 
are necessary:
\begin{eqnarray}
 _{N+2}\langle \phi_0 \mid \phi_0 \rangle_{N+2} \, = \, 2,  \ \ \
 _{N+2}\langle \phi_1 \mid \phi_1 \rangle_{N+2} \, = \, 2,  \ \ \
 _{N+1,N+2}\langle \phi_2 \mid \phi_2 \rangle_{N+1,N+2} \, = \, 4,  
\nonumber \\ 
 _{N+1,N+2}\langle \phi_3 \mid \phi_3 \rangle_{N+1,N+2} \, = \, 2,  \ \ \
 _{N+1,N+2}\langle \phi_4 \mid \phi_4 \rangle_{N+1,N+2} \, = \, 8,  \ \ \
 _{N+1,N+2}\langle \phi_5 \mid \phi_5 \rangle_{N+1,N+2} \, = \, 2,  \ \ \
\nonumber \\
 _{N,N+1,N+2}\langle \phi_6 \mid \phi_6 \rangle_{N,N+1,N+2} \, = \, 16, 
 \ \ \
 _{N,N+1,N+2}\langle \phi_7 \mid \phi_7 \rangle_{N,N+1,N+2} \, = \, 16,
\nonumber \\
 \langle N+1 \mid_{N+2} \langle \phi_0 \mid 
 \phi_2 \rangle_{N+1,N+2} \mid N \rangle \, = \, 
 (-1) \langle N+1 \mid \phi_0 \rangle_{N+1} \mid N \rangle ,
\nonumber \\
 \langle N+1 \mid_{N+2} \langle \phi_1 \mid 
 \phi_5 \rangle_{N+1,N+2} \mid N \rangle \, = \, 
 \langle N+1 \mid \phi_1 \rangle_{N+1} \mid N \rangle ,
\nonumber \\
 \langle N+1 \mid_{N+2} \langle \phi_0 \mid 
 \phi_6 \rangle_{N,N+1,N+2} \mid N-1 \rangle \, = \, 
 (-1) \langle N+1 \mid \phi_4 \rangle_{N,N+1} \mid N-1 \rangle ,
\nonumber \\
 \langle N \mid_{N+1,N+2} \langle \phi_4 \mid 
 \phi_7 \rangle_{N,N+1,N+2} \mid N-1 \rangle \, = \, 
 (-4) \langle N \mid \phi_0 \rangle_{N} \mid N-1 \rangle \, .
\end{eqnarray}
Using these inner products, a straightforward calculation gives
the following (coupled) recursion relations:
\begin{eqnarray}
Z_{N+2} \ & = & \ (2+2\alpha^2)Z_{N+1} \, - \, 2\beta Y_{N+1} 
          \, + \, 2\alpha \eta X_{N+1} 
	  \, + \, (4\beta^2 + 2\gamma^2 + 8\xi^2 + 2\eta^2)Z_N 
             \nonumber \\
	  & + & \, (16\delta^2 + 16\varepsilon^2 - 16\xi \delta)Z_{N-1}
          \, - \, 8\xi \varepsilon Y_N 
          \, + \, 8\delta \varepsilon Y_{N-1}  \, ,
 \nonumber \\
Y_{N+2} \ & = & \ 2Z_{N+1} \, - \, \beta Y_{N+1} 
          \, - \, 8\xi \delta Z_{N-1} 
          \, + \, 4\delta \varepsilon Y_{N-1}  \, ,
 \nonumber \\
X_{N+2} \ & = & \ 2\alpha Z_{N+1} \, + \, \eta X_{N+1}  \, ,
 \nonumber \\
E_{N+2} \ & = & \ (2+2\alpha^2)E_{N+1} 
          \, + \, (-8t_{\perp}\alpha + 2U\alpha^2)Z_{N+1}
          \, - \, 2\beta D_{N+1} \, + \, 2\alpha \eta C_{N+1}
 \nonumber \\
          & + & \, (8t\xi + 4t_{\perp}\alpha \beta)Y_{N+1}
	  \, + \, (-4t_{\perp}\eta - 8t\alpha \xi + 2U\alpha \eta)X_{N+1}
 \nonumber \\         
          & + & (4\beta^2 + 2\gamma^2 + 8\xi^2 + 2\eta^2)E_N
          \, + \, (-32t\beta \xi + 4U\gamma^2 - 16t\gamma \xi
                   +8U\xi^2 - 16t\xi \eta + 4U\eta^2)Z_N
 \nonumber \\
          & + & \, (16t\beta \delta - 8t\varepsilon 
                    + 8t\alpha^2 \varepsilon + 16t\beta \varepsilon
                    + 8t\gamma \varepsilon - 8U\xi \varepsilon
                    + 8t\eta \varepsilon - 8t\delta)Y_N
          \, + \, 8t\alpha \delta X_N
 \nonumber \\
          & + & \, (32t\beta \varepsilon + 16U\delta^2 
                    +16U\varepsilon^2 + 32t\beta \delta
                    +16t\gamma \delta - 16U\xi \delta
		    +16t\eta \delta 
                    + 32t_{\perp}\alpha \delta \xi)X_{N-1}
          \, - \, 8\xi \varepsilon D_N
 \nonumber \\
          & + & \, (16\delta^2 + 16\varepsilon^2 - 16\delta \xi)E_{N-1}
          \, + \, 8\delta \varepsilon D_{N-1}
          \, + \, (8U\delta \varepsilon 
 		   -16t_{\perp}\alpha \delta \varepsilon)Y_{N-1} \, ,
\nonumber \\
D_{N+2} \ & = & \ 2E_{N+1} \, - \, 4t_{\perp}\alpha Z_{N+1}
          \, - \, \beta D_{N+1} \, + \, 4t\xi Y_{N+1} 
	  \, - \, 2t_{\perp}\eta X_{N+1} 
	  \, + \, (-4t\varepsilon - 4t\delta)Y_N
	  \, + \, 4t\alpha \delta X_N
 \nonumber \\
          & + & \, (16t\beta \delta + 8t\gamma \delta
		    -8U\xi \delta + 8t\eta \delta 
		    +16t\beta \varepsilon)Z_{N-1}
          \, - \, 8\xi \delta E_{N-1} 
	  \, + \, 4\delta \varepsilon D_{n-1}
          \, + \, 4U\delta \varepsilon Y_{N-1}   \, ,
 \nonumber \\
C_{N+2} \ & = & \ 2\alpha E_{N+1} 
          \, + \, (-4t_{\perp} + 2U\alpha)Z_{N+1}
          \, + \, 2t_{\perp}\beta Y_{N+1}
	  \, + \, (-4t\xi + U\eta)X_{N+1} \, + \, \eta C_{N+1}
 \nonumber \\
	  & + & \, 16t_{\perp}\xi \delta Z_{N-1}
	  \, - \, 8t_{\perp}\delta \varepsilon Y_{N-1}
	  \, + \, 4t\alpha \varepsilon Y_N  \, .
\end{eqnarray}


\section{The Doped Ladder}
For the doped ladder (see Eq.~\ref{dopedansatz}), 
 $\mid \phi_0 \rangle_{N+2}$ ,$\mid \phi_1 \rangle_{N+2}$,
$\mid \phi_2 \rangle_{N+1,N+2}$ ,$\mid \phi_3 \rangle_{N+1,N+2}$,
$\mid \phi_4 \rangle_{N+1,N+2}$ ,$\mid \phi_5 \rangle_{N+1,N+2}$,
$\mid \phi_6 \rangle_{N,N+1,N+2}$, and $\mid \phi_7 \rangle_{N,N+1,N+2}$
are the same as the half-filled case, and
\begin{eqnarray}
 \mid \phi_0^h \rangle_{N+2} & = & \mid 0 \rangle_{N+2} \, , 
 \nonumber \\
 \mid \phi_1 \rangle_{N+1,N+2} & = & 
     \Delta^{\dagger}_{(N+1,1),(N+2,2)} \mid 0 \rangle_{N+1,N+2} + 
     \Delta^{\dagger}_{(N+2,1),(N+1,2)} \mid 0 \rangle_{N+1,N+2} \, ,
 \nonumber \\
 \mid \phi_2 \rangle_{N+1,N+2} & = &
     \Delta^{\dagger}_{(N+1,1),(N+2,1)} \mid 0 \rangle_{N+1,N+2} + 
     \Delta^{\dagger}_{(N+1,2),(N+2,2)} \mid 0 \rangle_{N+1,N+2} \, ,
 \nonumber \\
 \mid \phi_3 \rangle_{N,N+1,N+2} & = &
     \Delta^{\dagger}_{(N,1),(N+2,1)} \Delta^{\dagger}_{(N+1,2),(N+2,2)}
     \mid 0 \rangle_{N,N+1,N+2} + 
     \Delta^{\dagger}_{(N,2),(N+2,2)} \Delta^{\dagger}_{(N+1,1),(N+2,1)}
     \mid 0 \rangle_{N,N+1,N+2}  \, ,
 \nonumber \\
 \mid \phi_4 \rangle_{N,N+1,N+2} & = &
     \Delta^{\dagger}_{(N,1),(N+2,1)} \Delta^{\dagger}_{(N,2),(N+1,2)}
     \mid 0 \rangle_{N,N+1,N+2} + 
     \Delta^{\dagger}_{(N,2),(N+2,2)} \Delta^{\dagger}_{(N,1),(N+1,1)}
     \mid 0 \rangle_{N,N+1,N+2} \, .
\end{eqnarray}

To derive the recursion relations, we use the inner products from the 
half-filled case as well as the the following:
\begin{eqnarray}
 _{N+2}\langle \phi_0^h \mid \phi_0^h \rangle_{N+2} \, = \, 1,  
  \ \ \
 _{N+1,N+2}\langle \phi_1^h \mid \phi_1^h \rangle_{N+1,N+2} \, = \, 4,
  \ \ \
 _{N+1,N+2}\langle \phi_2^h \mid \phi_2^h \rangle_{N+1,N+2} \, = \, 4,  
\nonumber \\ 
 _{N,N+1,N+2}\langle \phi_3^h \mid \phi_3^h \rangle_{N,N+1,N+2} \, = \, 8,  
  \ \ \
 _{N,N+1,N+2}\langle \phi_4^h \mid \phi_4^h \rangle_{N,N+1,N+2} \, = \, 8,  
\nonumber \\
 \langle N+1 P+1\mid_{N+2} \langle \phi_0 \mid 
 \phi_3^h \rangle_{N,N+1,N+2} \mid N-1 P \rangle \, = \, 
 (-1) \langle N+1 P+1\mid \phi_1^h \rangle_{N,N+1} \mid N-1 P \rangle ,
\nonumber \\
 \langle N P \mid_{N+1,N+2} \langle \phi_1^h \mid 
 \phi_4^h \rangle_{N,N+1,N+2} \mid N-1 P \rangle \, = \, 
 (-2) \langle N P \mid \phi_0 \rangle_{N} \mid N-1 P \rangle  \, .
\end{eqnarray}
Using these inner products, a straightforward calculation gives the 
following (coupled) recursion relations:
\begin{eqnarray}
Z_{N+2,P+1} \ & = & \ (2+2\alpha^2)Z_{N+1,P+1} \, - \, 2\beta Y_{N+1,P+1} 
          \, + \, 2\alpha \eta X_{N+1,P+1} 
	  \, + \, (4\beta^2 + 2\gamma^2 + 8\xi^2 + 2\eta^2)Z_{N,P+1} 
             \nonumber \\
	  & + & \, (16\delta^2 + 16\varepsilon^2 - 16\xi \delta)Z_{N-1,P+1}
          \, - \, 8\xi \varepsilon Y_{N,P+1} 
          \, + \, 8\delta \varepsilon Y_{N-1,P+1}    	     
	  \, + \, \omega^2 Z_{N+1,P} 
	  \, + \, (4\lambda^2 + 4\zeta^2)Z_{N,P}
	     \nonumber \\
          & - & \, 4\lambda\nu Y_{N,P}
	  \, + \, (8\mu^2 + 8\nu^2 - 8\lambda\mu)Z_{N-1,P}
	  \, + \, 4\mu\nu Y_{N-1,P}  \, ,	  
 \nonumber \\
Y_{N+2,P+1} \ & = & \ 2Z_{N+1,P+1} \, - \, \beta Y_{N+1,P+1} 
          \, - \, 8\xi \delta Z_{N-1,P+1} 
          \, + \, 4\delta \varepsilon Y_{N-1,P+1}
	  \, - \, 4\mu \lambda Z_{N-1,P}
	  \, + \, 2\mu \nu Y_{N-1,P}  \, ,
 \nonumber \\
X_{N+2,P+1} \ & = & \ 2\alpha Z_{N+1,P+1} \, + \, \eta X_{N+1,P+1} \, ,
 \nonumber \\
E_{N+2,P+1} \ & = & \ (2+2\alpha^2)E_{N+1,P+1} 
          \, + \, (-8t_{\perp}\alpha + 2U\alpha^2)Z_{N+1,P+1}
          \, - \, 2\beta D_{N+1,P+1} \, + \, 2\alpha \eta C_{N+1,P+1}
 \nonumber \\
         & + & \, (8t\xi + 4t_{\perp}\alpha \beta)Y_{N+1,P+1}
	  \, + \, (-4t_{\perp}\eta - 8t\alpha \xi + 2U\alpha \eta)X_{N+1,P+1}
 \nonumber \\         
         & + & \, (4\beta^2 + 2\gamma^2 + 8\xi^2 + 2\eta^2)E_{N,P+1}
          \, + \, (-32t\beta \xi + 4U\gamma^2 - 16t\gamma \xi
                   +8U\xi^2 - 16t\xi \eta + 4U\eta^2)Z_{N,P+1}
 \nonumber \\
         & + & \, (16t\beta \delta - 8t\varepsilon 
                   + 8t\alpha^2 \varepsilon + 16t\beta \varepsilon
                   + 8t\gamma \varepsilon - 8U\xi \varepsilon
                   + 8t\eta \varepsilon - 8t\delta)Y_{N,P+1}
          \, + \, 8t\alpha \delta X_{N,P+1}
 \nonumber \\
         & + & \, (32t\beta \varepsilon + 16U\delta^2 
                    +16U\varepsilon^2 + 32t\beta \delta
                    +16t\gamma \delta - 16U\xi \delta
		    +16t\eta \delta 
                    + 32t_{\perp}\alpha \delta \xi)X_{N-1,P+1}
          \, - \, 8\xi \varepsilon D_{N,P+1}
 \nonumber \\
         & + & \, (16\delta^2 + 16\varepsilon^2 - 16\delta \xi)E_{N-1,P+1}
          \, + \, 8\delta \varepsilon D_{N-1,P+1}
          \, + \, (8U\delta \varepsilon 
 		   -16t_{\perp}\alpha \delta \varepsilon)Y_{N-1,P+1}
 \nonumber \\
        & + & \,  E_{N+1,P} 
         \, + \, (4\lambda^2 + 4\zeta^2) E_{N,P}
	 \, + \, (8\mu^2 + 8\nu^2 - 8\lambda \mu)E_{N-1,P}
	 \, - \, 4t \lambda Y_{N+1,P}
         \, - \, 4t \zeta X_{N+1,P}
 \nonumber \\
        & + & \, (-16t_{\perp}\lambda \zeta - 8t\lambda 
                  - 8t\alpha \zeta)Z_{N,P}
         \, + \, (4t\nu + 8t_{\perp}\zeta \nu
                  + 4t \mu + 4t \nu)Y_{N,P}
 \nonumber \\
          & + & (-16t\beta \nu + 8t \mu 
		 +16t_{\perp}\zeta \mu + 16t_{\perp}\alpha \lambda \mu
		 -16t\beta \mu)Z_{N-1,P}
	 \, - \, 4\lambda \nu D_{N,P}
	 \, + \, 4\mu \nu D_{N-1,P}
 \nonumber \\
	& - & \, 8t_{\perp} \alpha \mu \nu Y_{N-1,P}  \, ,
 \nonumber \\
D_{N+2,P+1} \ & = & \ 2E_{N+1,P+1} \, - \, 4t_{\perp}\alpha Z_{N+1,P+1}
          \, - \, \beta D_{N+1,P+1} \, + \, 4t\xi Y_{N+1,P+1} 
	  \, - \, 2t_{\perp}\eta X_{N+1,P+1}
 \nonumber \\        
         & + & \, (-4t\varepsilon - 4t\delta)Y_{N,P+1}
	  \, + \, 4t\alpha \delta X_{N,P+1}
          \, + \, (16t\beta \delta + 8t\gamma \delta
		    -8U\xi \delta + 8t\eta \delta 
		    +16t\beta \varepsilon)Z_{N-1,P+1}
 \nonumber \\     
         & - & \, 8\xi \delta E_{N-1,P+1} 
	  \, + \, 4\delta \varepsilon D_{n-1,P+1}
          \, + \, 4U\delta \varepsilon Y_{N-1,P+1}
         \, - \, 4t \lambda Z_{N,P}
	 \, + \, (2t \nu + 2t \mu)Y_{N,P}
 \nonumber \\	
         & - & \, 4\lambda \mu E_{N-1,P}
	 \, + \, 2\mu \nu D_{N-1,P}
	 \, + \, (4t \mu + 8t_{\perp} \zeta \mu)Z_{N-1,P} \, ,
 \nonumber \\
C_{N+2,P+1} \ & = & \ 2\alpha E_{N+1,P+1} 
          \, + \, (-4t_{\perp} + 2U\alpha)Z_{N+1,P+1}
          \, + \, 2t_{\perp}\beta Y_{N+1,P+1}
	  \, + \, (-4t\xi + U\eta)X_{N+1,P+1} 
 \nonumber \\
         & + & \, \eta C_{N+1,P+1}
	  \, + \, 16t_{\perp}\xi \delta Z_{N-1,P+1}
	  \, - \, 8t_{\perp}\delta \varepsilon Y_{N-1,P+1}
	  \, + \, 4t\alpha \varepsilon Y_{N,P+1}  
	  \, - \, 4t \zeta Z_{N,P}
 \nonumber \\
	 & + & \, 8t_{\perp} \lambda \mu Z_{N-1,P}
	  \, - \, 4t_{\perp} \mu \nu Y_{N-1,P}   \, .
\end{eqnarray}
\breakoff


\end{multicols}

\begin{thebibliography}{3}
\bibitem{tensorproduct}
Note that 
 $\mid \phi_i \rangle_{N+2,N+1} \mid N \rangle $ should (more precisely)
be written $\mid \phi_i \rangle_{N+2,N+1} \otimes \mid N \rangle $.  
However, throughout the paper, to make things less notationally cumbersome, 
we have omitted the tensor product symbol.
\end{thebibliography}
\end{document}